\documentclass[12pt]{article}
\usepackage{epsfig,epsf}
\def\beq{\begin{equation}}
\def\eeq{\end{equation}}
\def\eq#1{{Eq.~(\ref{#1})}}

\begin{document}
\begin{flushright}
BNL--NT--01/18\\
nucl-th/0108006\\
August 2, 2001

\end{flushright}

\begin{center}
{\large\bf Manifestations of High Density QCD in the first RHIC data}
\vskip1cm

Dmitri Kharzeev${}^{a)}$  and Eugene Levin${}^{b)}$
\vskip1cm

{\it a) Physics Department,}\\
{\it Brookhaven National Laboratory,}\\
{\it Upton, NY 11973 - 5000, USA}\\
\vskip0.3cm
{\it b ) HEP Department, School of Physics,}\\
{\it Raymond and Beverly Sackler Faculty of Exact Science,}\\
{\it Tel Aviv University, Tel Aviv 69978, Israel}\\

\end{center}
\bigskip
\begin{abstract}
We derive a simple analytical scaling function which embodies the 
predictions of high density QCD on
 the energy, centrality, rapidity, and atomic number dependences of hadron multiplicities 
in nuclear collisions.
Both centrality and rapidity dependences  
of hadron multiplicity in $Au-Au$ collisions as measured at RHIC at $\sqrt{s}= 130 \ {\rm GeV}$ 
are well described in this approach. 
The centrality and rapidity dependences of hadron 
multiplicity at $\sqrt{s}= 200 \ {\rm GeV}$ run at RHIC are predicted; 
the variation of these dependences with energy appear different from 
other approaches, and can be used as an important test of the ideas 
based on parton saturation and classical Chromo-Dynamics.  

\end{abstract}
\vskip0.3cm


Relativistic heavy ion collisions allow to probe QCD in the non--linear 
regime of high parton density and high color field strength. 
It has been conjectured long time ago that the dynamics of QCD  
in the high density domain  may become qualitatively different: in parton language, 
this is best described in terms of {\it parton saturation} \cite{GLR,MUQI,BM}, and in the language 
of color fields -- in terms of the {\it classical} Chromo--Dynamics \cite{MV,YM,KV,KN}. 
In this high density regime,  
the transition amplitudes are dominated not by quantum fluctuations, but by 
the configurations of classical field containing large, $\sim 1/\alpha_s$, 
numbers of gluons. One thus uncovers new 
non--linear features of QCD, 
which cannot be investigated in the more traditional applications
based on the perturbative approach. 
The classical color fields in the initial nuclei (the 
``color glass condensate'' \cite{MV,YM}) can be thought of as 
either perturbatively generated, or 
as being a topologically non--trivial superposition of the Weizs{\"a}cker-Williams 
radiation and the quasi--classical vacuum fields \cite{inst,inst1}.    
 
\vskip0.3cm

Recently, with the advent of RHIC, the heavy ion program has entered the  
era of collider experiments. 
First RHIC results on hadron multiplicities have been presented \cite{data1,data2,data3,data4}. 
The established experimental centrality dependence is so far in accord with 
predictions \cite{KN} based on high density QCD. (For alternative approaches, 
see \cite{WG,EKRT,Alf,alt}). 
Rapidity dependence 
of the hadron multiplicity has recently been presented as well, and  
it is important to check if it agrees with the high density QCD calculations as well. 
It is also most timely to make predictions for the higher energy, $\sqrt{s} = 200 \ 
{\rm GeV}$ run. These are the objectives of this paper.

\vskip0.3cm

Let us begin by presenting a brief and elementary introduction into 
the concept of parton saturation \cite{GLR}. 
Consider an external probe $J$ interacting with the 
nuclear target of atomic number $A$ (see Fig. 1). 
At small values of Bjorken $x$, 
by uncertainty principle the interaction develops over large 
longitudinal distances $z \sim 1/(mx)$, where $m$ is the 
nucleon mass. As soon as $z$ becomes larger than the nuclear diameter, 
the probe cannot distinguish between the nucleons located on the front and back edges 
of the nucleus, and all partons within the transverse area $\sim 1/Q^2$ 
determined 
by the momentum transfer $Q$ participate in the interaction coherently. 
The density of partons in the transverse plane is given by
\beq
\rho_A \simeq {x G_A(x,Q^2) \over \pi R_A^2} \sim A^{1/3},
\eeq
where we have assumed that the nuclear gluon distribution  
scales with the number of nucleons $A$. The probe interacts with 
partons with cross section $\sigma \sim \alpha_s / Q^2$; therefore, 
depending on the magnitude of momentum transfer $Q$, atomic number $A$, 
and the value of Bjorken $x$, one may encounter two regimes:
\begin{itemize}
\item{$\sigma \rho_A \ll 1$ -- this is a familiar ``dilute'' regime of 
incoherent interactions, which is well described by the methods of 
perturbative QCD;}
\item{$\sigma \rho_A \gg 1$ -- in this regime, we deal with a dense 
parton system. Not only do the ``leading twist'' expressions become 
inadequate, but also the expansion in higher twists, i.e. in 
multi--parton correlations, breaks down here.}
\end{itemize}

The border between the two regimes can be found from the condition 
$\sigma \rho_A \simeq 1$; it determines the critical value of the 
momentum transfer (``saturation scale''\cite{GLR,MV}) at which the parton system 
becomes to look dense to the probe\footnote{Note that since 
$Q_s^2 \sim A^{1/3}$, this expression 
in the target rest frame can also be understood as describing a broadening 
of the transverse momentum resulting from the multiple re-scattering 
of the probe.}:
\beq 
Q_s^2 \sim  \alpha_s \ {x G_A(x,Q_s^2) \over \pi R_A^2}. \label{qsat}
\eeq
In this regime, the number of gluons from (\ref{qsat}) is given by 
\beq
x G_A(x,Q_s^2) \sim {\pi \over \alpha_s(Q_s^2)}\ Q_s^2 R_A^2, \label{gsat}
\eeq
where $Q_s^2 R_A^2 \sim A$. 
One can see that the number of gluons 
is proportional to the {\em inverse} of $\alpha_s(Q_s^2)$, and 
becomes large in the weak coupling regime. In this regime,
as we shall now discuss, the dynamics is likely to become 
essentially classical. 
\begin{figure}[h]
\begin{minipage}{10cm}

\begin{center}   
\epsfysize=6.4cm
\leavevmode
\hbox{ \epsffile{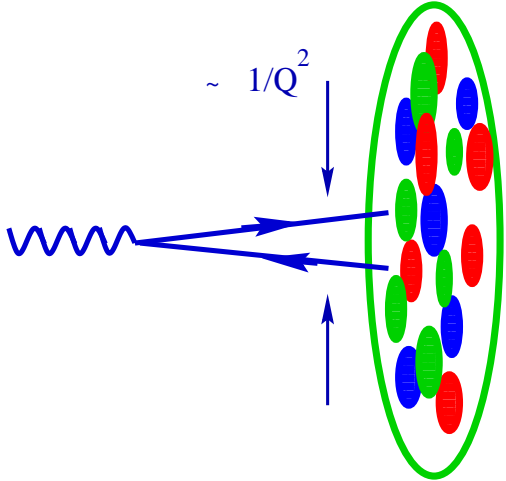}}
\end{center}  
\end{minipage}
\begin{minipage}{5.5cm} 
\caption{Hard probe interacting with the nuclear target 
resolves the transverse area $\sim 1/Q^2$ ($Q^2$ is the square of the 
momentum transfer) and, in the target rest frame, the longitudinal 
distance $\sim 1/(m x)$ ($m$ is the nucleon mass and $x$ -- Bjorken variable).} 
\label{fig:satpict}
\end{minipage}
\end{figure}

\vskip0.3cm

The condition (\ref{qsat}) can be re--derived in a way \cite{DK} which illustrates the link 
between the saturation and classical Yang--Mills fields \cite{MV}. 
As a first step, let us re-scale the gluon fields 
in the Lagrangian 
\beq
{\cal{L}} = -{1 \over 4} G_{\mu\nu}^a G_{\mu\nu}^a + \sum_f \bar{q}_f^a 
(i \gamma_{\mu} D_{\mu} - m_f) q_f^a; \label{lagr}
\eeq
as follows: $A_{\mu}^a \to \tilde{A}_{\mu}^a = 
g A_{\mu}^a$. In terms of new fields, $\tilde{G}_{\mu \nu}^a = 
g G_{\mu \nu}^a = \partial_{\mu} \tilde{A}_{\nu}^a - \partial_{\nu} 
\tilde{A}_{\mu}^a +  f^{abc} \tilde{A}_{\mu}^b \tilde{A}_{\nu}^c$, 
and the dependence of the action corresponding to the 
Lagrangian (\ref{lagr}) on the coupling constant is given by  
\beq
S \sim \int {1 \over g^2}\ \tilde{G}_{\mu \nu}^a  \tilde{G}_{\mu \nu}^a 
\ d^4 x. \label{act}
\eeq
Let us now consider a classical configuration of gluon fields; by definition, 
$\tilde{G}_{\mu \nu}^a$ in such a configuration does not depend on 
the coupling, and the action is large, $S \gg \hbar$. The number of 
quanta in such a configuration is then
\beq
N_g \sim {S \over \hbar} \sim {1 \over \alpha_s}\ \rho_4 V_4, \label{numb}
\eeq
where we re-wrote (\ref{act}) as a product of four--dimensional 
action density $\rho_4$ and the four--dimensional volume $V_4$. 
 
The effects of non--linear interactions among the gluons become 
important when $\partial_{\mu} \tilde{A}_{\mu} \sim \tilde{A}_{\mu}^2$ 
(this condition can be made explicitly gauge invariant if we derive it 
from the expansion of a correlation function of gauge-invariant 
gluon operators, e.g., $\tilde{G}^2$). In momentum space, this 
equality corresponds to 
\beq
Q_s^2 \sim \tilde{A}^2 \sim (\tilde{G}^2)^{1/2} = 
\sqrt{\rho_4}; \label{nonlin}
\eeq
$Q_s$ is the typical value of the gluon momentum below which 
the interactions become essentially non--linear. 

Consider now a nucleus $A$ boosted to a high momentum. By uncertainty 
principle, the gluons with transverse momentum $Q_s$ are extended 
in the longitudinal and proper time directions by $\sim 1/Q_s$; 
since the transverse area is $\pi R_A^2$, the four--volume 
is $V_4 \sim \pi R_A^2 / Q_s^2$. The resulting four--density from 
(\ref{numb}) is then 
\beq
\rho_4 \sim \alpha_s\ {N_g \over V_4} \sim \alpha_s\ {N_g\ Q_s^2 
\over \pi R_A^2} 
\sim Q_s^4, \label{class}
\eeq
where at the last stage we have used the non--linearity condition (\ref{nonlin}),  
$\rho_4 \sim Q_s^4$. It is easy to see that (\ref{class}) coincides with the 
saturation condition (\ref{qsat}), since the number of gluons in the 
infinite momentum frame $N_g \sim x G(x,Q_s^2)$. This simple derivation  
illustrates that the physics in the high--density regime can potentially 
be understood in terms of classical gluon fields. 
This correspondence allowed to formulate an effective 
quasi--classical theory \cite{MV}, which is a subject of vigorous  
investigations at present (see, e.g., \cite{YM,KV}).

\vskip0.3cm

In nuclear collisions, the saturation scale becomes a function of centrality; 
a generic feature of the quasi--classical 
approach -- the proportionality of the number of gluons to the inverse 
of the coupling constant (\ref{numb}) -- thus leads to definite predictions \cite{KN} 
on the centrality dependence of multiplicity:
\beq
{dN \over d \eta} = c\ N_{part} \ xG(x, Q_s^2), \label{mults1}
\eeq
where $xG(x, Q_s^2) \sim 1/\alpha_s(Q_s^2) \sim \ln(Q_s^2/\Lambda_{QCD}^2)$, 
$N_{part}$ is the number of participants, and $c$ is the ``parton liberation'' coefficient, 
which was determined \cite{KN} from the experimental multiplicity to be 
\beq
c = 1.23 \pm 0.20. \label{val}
\eeq
This value is on the order of unity \cite{AHM}, and 
agrees with the lattice \cite{Raju} and analytical \cite{Yuri} calculations.  
The prediction (\ref{mults1}) so far is 
in accord with the data coming from RHIC \cite{data1,data2,data3,data4}. Let us now make predictions 
for $\sqrt{s} = 200\ \rm{GeV}$ collisions based on this picture. 

\vskip0.3cm

The energy dependence of the hadron production is determined by the variation of saturation scale 
$Q_s$ with Bjorken $x = Q_s / \sqrt{s}$. This variation, in turn, is determined by the 
$x-$ dependence of the gluon structure function. 
In the saturation approach, the gluon distribution is related to the saturation scale 
by Eq.(\ref{qsat}). 
A good description of HERA data is obtained with saturation scale $Q_s^2 = 1 \div 2\ \rm{GeV}^2$
with $W$ - dependence ($W \equiv \sqrt{s}$ is the center-of-mass energy available 
in the photon--nucleon system)  \cite{GW} 
\beq
Q^2_s \,\,\propto\, W^{\lambda},
\eeq
where $\lambda \simeq 0.25 \div 0.3$. In spite of significant uncertainties in the determination 
of the gluon structure functions, perhaps even more important is the observation \cite{GW} that the 
HERA data exhibit scaling when plotted as a function of variable 
\beq
\tau \,=\, {Q^2 \over Q_0^2} \ \left({x \over x_0}\right)^{\lambda}, 
\eeq
where the value of $\lambda$ is again within the limits $\lambda \simeq 0.25 \div 0.3$. 
In high density QCD, this scaling is a consequence 
of the existence of dimensionful scale \cite{GLR,MV}) 
\beq
Q_s^2(x) = Q_0^2 \ (x_0 / x)^{\lambda}. 
\eeq
The formula (\ref{mults1}) can be equivalently re--written as (see \cite{KN} 
for details)
\beq
{dN \over d \eta} \sim S_A \ {Q_s^2 \over \alpha_s(Q_s^2)}, \label{mults}
\eeq
where $S_A$ is the nuclear overlap area, 
determined by atomic number and the centrality of collision.

The energy dependence of multiplicity is thus (up to a logarithmic correction, which is small 
for the energy interpolation that we make) given by 
\beq
{dN \over d \eta} (\sqrt{s} = 200\ \rm{GeV}) \simeq \left({200 \over 130}\right)^{\lambda} \ 
{dN \over d \eta} (\sqrt{s} = 130\ \rm{GeV}) = (1.10 \div 1.14)\ {dN \over d \eta} (\sqrt{s} 
= 130\ \rm{GeV}), 
\eeq
indicating that the hadron multiplicity is expected to raise in the $\sqrt{s} = 200\ \rm{GeV}$ 
run by about $10\div14 \%$\footnote{At the time when this paper is being finalized, 
the result of the PHOBOS 
multiplicity measurement for central collisions has just been announced -- see Note added.}. 
Taking the PHOBOS number for charged multiplicity at $\sqrt{s} = 130\ \rm{GeV}$ 
for the $6 \%$ centrality cut, 
$dN/d\eta = 555 \pm 12(stat) \pm 35(syst)$ \cite{PHOBOS130}, one gets a prediction of 
\beq
dN/d\eta = 616 \div 634 \label{pred200}
\eeq
 for the $\sqrt{s} = 200\ \rm{GeV}$ (these are the central values not 
taking into 
account the error bars of the $\sqrt{s} = 130\ \rm{GeV}$ measurement). 
 
One can also try to extract the value of $\lambda$ from the energy dependence of hadron 
multiplicity measured by PHOBOS at $\sqrt{s} = 130\ \rm{GeV}$ and at at $\sqrt{s} = 56\ \rm{GeV}$; 
this procedure yields $\lambda \simeq 0.37$, which is larger than 
the value inferred from the HERA data (and is very close to the value $\lambda \simeq 0.38$, 
resulting from the final--state saturation calculations \cite{EKRT}).

\vskip0.3cm

Let us now proceed to the calculation of the (pseudo)rapidity and centrality dependences. 
We need to evaluate the leading tree diagram describing 
emission of gluons on the classical level, see Fig. 2. 
\begin{figure}[h]
\begin{minipage}{9.5cm}   
\begin{center}
\epsfysize=9.4cm
\leavevmode
\hbox{ \epsffile{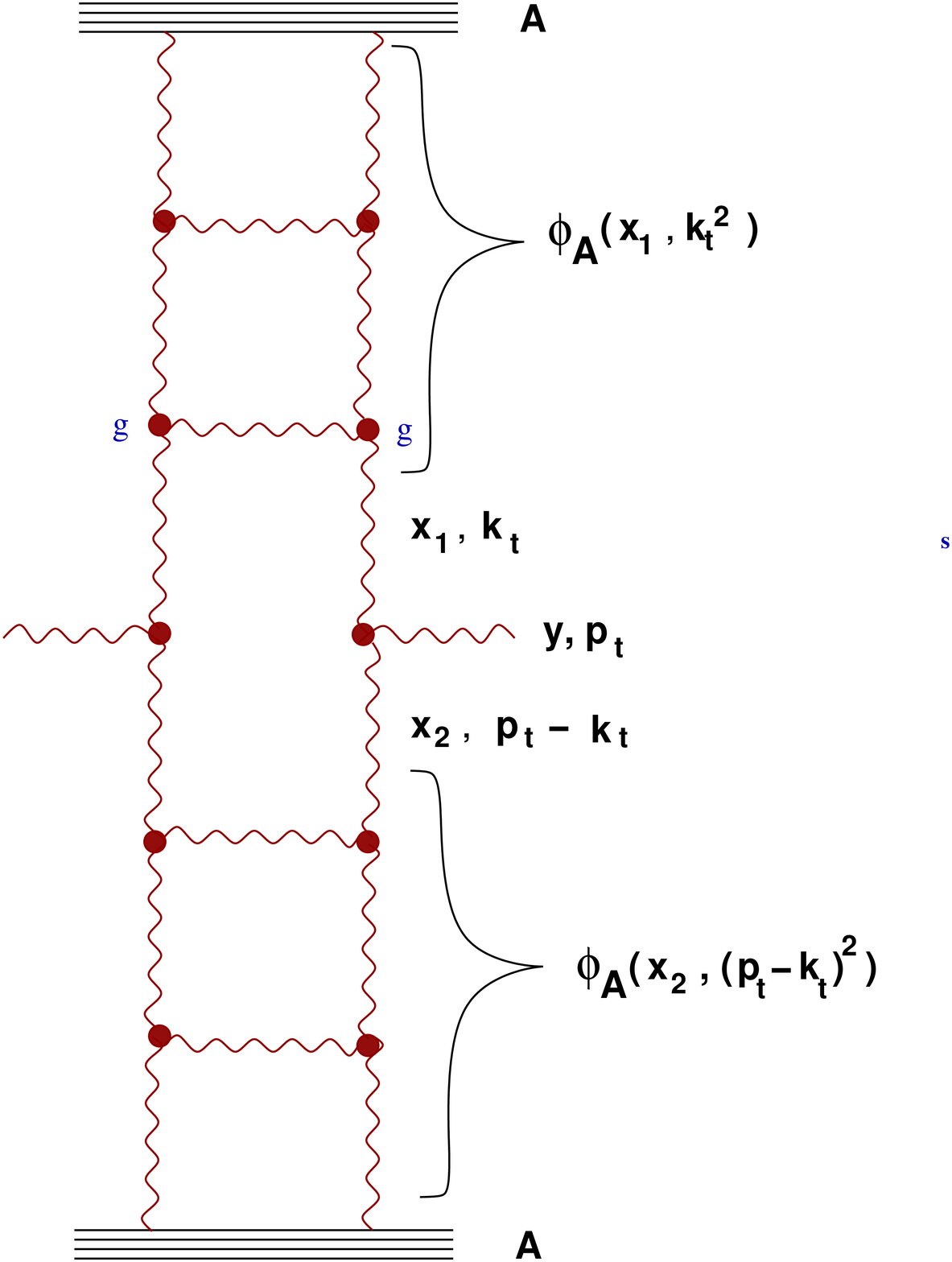}}
\end{center}
\end{minipage}
\begin{minipage}{5.5cm}
\caption{ The Mueller diagram for the classical gluon radiation.}
\label{phi}
\end{minipage}
\end{figure}
Let us introduce the unintegrated gluon distribution $\varphi_A (x, k_t^2)$ which 
describes the probability to find a gluon with a given $x$ and transverse 
momentum $k_t$ inside the nucleus $A$. As follows from this definition, 
the unintegrated distribution is related to the gluon structure function by
\beq
xG_A(x, p_t^2) = \int^{p_t^2} d k_t^2 \ \varphi_A(x, k_t^2);
\eeq
when $p_t^2 > Q_s^2$, the unintegrated distribution corresponding to the bremsstrahlung 
radiation spectrum is 
\beq
\varphi_A(x, k_t^2) \sim {\alpha_s \over \pi} \ {1 \over k_t^2}.
\eeq 
In the saturation region, the gluon structure function is given by 
(\ref{gsat}); the corresponding unintegrated gluon distribution has only logarithmic dependence on the 
transverse momentum: 
\beq
\varphi_A(x, k_t^2) \sim {S_A \over \alpha_s}; \ k_t^2 \leq Q_s^2, \label{unint}
\eeq
where $S_A$ is the nuclear overlap area, determined by the atomic numbers of the 
colliding nuclei and by centrality of the collision.

 The differential cross section 
of gluon production in a $AA$ collision can now be written down as \cite{GLR,GM}
\beq
E {d \sigma \over d^3 p} = {4 \pi N_c \over N_c^2 - 1}\ {1 \over p_t^2}\ \int d k_t^2 \ 
\alpha_s \ \varphi_A(x_1, k_t^2)\ \varphi_A(x_2, (p-k)_t^2), \label{gencross}    
\eeq
where $x_{1,2} = (p_t/\sqrt{s}) \exp(\pm \eta)$, with $\eta$ the (pseudo)rapidity of the 
produced gluon; the running coupling $\alpha_s$ has to be evaluated at the 
scale $Q^2 = max\{k_t^2, (p-k)_t^2\}$. 
The rapidity density is then evaluated from (\ref{gencross}) according to 
\beq
{dN \over d y} = {1 \over \sigma_{AA}}\ \int d^2 p_t \left(E {d \sigma \over d^3 p}\right), 
\label{rapden}
\eeq
where $\sigma_{AA}$ is the inelastic cross section of nucleus--nucleus interaction.

Since the rapidity $y$ and Bjorken variable are related by $\ln 1/x = y$, 
the $x-$ dependence of the gluon structure function translates into the following 
dependence of the saturation scale $Q_s^2$ on rapidity:
\beq
Q_s^2(s; \pm y) = Q_s^2(s; y = 0)\ \exp(\pm \lambda y). \label{qsy}
\eeq

As it follows from (\ref{qsy}), the increase of rapidity at a fixed $W \equiv \sqrt{s}$ 
moves the wave function of one of the colliding 
nuclei deeper into the saturation region, while leading to a 
smaller gluon density in the other, which as a result can be 
pushed out of the saturation domain. Therefore, depending on the value of rapidity, 
the integration over the transverse momentum in Eqs. (\ref{gencross}),(\ref{rapden}) can be split in 
two regions: i) the region $\Lambda_{QCD} < k_t < Q_{s,min}$ in which the wave 
functions are both 
in the saturation domain; and ii) the region  $\Lambda << Q_{s,min} < k_t < Q_{s,max}$ in which 
the wave function of 
one of the nuclei is in the saturation region and the other one is not. 
Of course, there is 
also the region of $k_t > Q_{s,max}$, which is governed by the usual perturbative dynamics, 
but our assumption here is that the r{\^o}le of these genuine hard processes in the bulk 
of gluon production is relatively small; in the saturation scenario, 
these processes represent quantum fluctuations above the classical background. It is worth 
commenting that in the conventional mini--jet picture, this classical background is absent, 
and the multi--particle production is dominated by perturbative processes. 
This is the main physical difference between the two approaches; for the production 
of particles with $p_t >> Q_s$ they lead to identical results.    

To perform the calculation according to (\ref{rapden}),(\ref{gencross}) away from $y=0$ we need also 
to specify the behavior of the gluon structure function at large Bjorken $x$ (and out of 
the saturation region).  
At $x \to 1$, this behavior is governed by the QCD counting rules \cite{qcr}, $xG(x) \sim (1-x)^4$, so 
we adopt the following conventional form: $xG(x) \sim x^{-\lambda}\ (1-x)^4$.
  
We now have everything at hand to perform the integration over transverse momentum 
in (\ref{rapden}), (\ref{gencross}); the result is the 
following:
\beq 
{dN \over d y} = const\ S_A\ Q_{s,min}^2 \ \ln\left({Q_{s,min}^2 \over \Lambda_{QCD}^2}\right) 
\left[1 + {1 \over 2}\ \ln\left({Q_{s,max}^2 \over Q_{s,min}^2}\right)\ 
\left(1 - {Q_{s,max} \over \sqrt{s}} e^{|y|}\right)^4\right],
 \label{resy}    
\eeq 
where the constant is energy--independent, $S_A$ is the nuclear overlap area, 
$Q_s^2 \equiv Q_s^2(s; y = 0)$, and $Q_{s,min(max)}$ 
are defined as the smaller (larger) values of (\ref{qsy}); at $y=0$, 
$Q_{s,min}^2 = Q_{s,max}^2 = Q_s^2(s) = Q_s^2(s_0)\ (s /s_0)^{\lambda / 2}$. 
The first term in the brackets in (\ref{resy}) originates from the region in which both 
nuclear wave functions are in the saturation regime; this corresponds to 
the familiar $\sim (1/\alpha_s)\ Q_s^2 R_A^2$ term in the gluon multiplicity, cf. Eq.(\ref{gsat}). 
The second term comes from the region in which only one of the wave functions is in 
the saturation region. The coefficient $1/2$ in front of the second term in 
square brackets comes from $k_t$ ordering of gluon momenta in evaluation of 
the integral of Eq.(\ref{gencross}).

The formula (\ref{resy}) has been derived using the form (\ref{unint}) 
for the unintegrated gluon distributions. We have checked numerically that the use of 
more sophisticated functional form of  
$\varphi_A$ taken from the saturation model of Golec-Biernat and W{\"u}sthoff \cite{GW} 
in Eq.(\ref{gencross}) affects the results only at the level of about $3\%$.

\vskip0.3cm

Since $S_A Q_s^2 \sim N_{part}$ (recall that $Q_s^2 \gg  \Lambda_{QCD}^2$ is defined 
as the density of partons 
in the transverse plane, which is proportional to the density of participants), we can 
re--write (\ref{resy}) in the following final form
\beq
{dN \over d y} = c\ N_{part}\ \left({s \over s_0}\right)^{\lambda \over 2}\ e^{- \lambda |y|}\ 
\left[\ln\left({Q_s^2 \over \Lambda_{QCD}^2}\right) - \lambda |y|\right]\ 
\left[ 1 +  \lambda |y| \left( 1 - {Q_s \over \sqrt{s}}\ e^{(1 + \lambda/2) |y|} \right)^4 \right],  
\label{finres}
\eeq
with $Q_s^2(s) = Q_s^2(s_0)\ (s /s_0)^{\lambda / 2}$.
This formula is the central result of our paper; it expresses the predictions of 
high density QCD for the energy, centrality, rapidity, and atomic number dependences 
of hadron multiplicities in nuclear collisions in terms of a single scaling function. 
Once the energy--independent constant $c \sim 1$ and $Q_s^2(s_0)$ are determined 
at some energy $s_0$, Eq. (\ref{finres}) contains no free parameters. (The value of $\lambda$, 
describing the growth 
of the gluon structure functions at small $x$ can be determined in deep--inelastic scattering; 
the HERA data are fitted with $\lambda \simeq 0.25 \div 0.3$ \cite{GW}).
At $y = 0$ the expression (\ref{resy}) coincides exactly with the one 
derived in \cite{KN}, and extends it to describe the rapidity and energy dependences.

Before we can compare (\ref{resy}) to the data, we have to take account of the 
difference between rapidity $y$ and the measured pseudo-rapidity $\eta$. This is done 
by multiplying (\ref{resy}) by the Jacobian of the $y \leftrightarrow \eta$ transformation;
a simple calculation yields
\beq    
h(\eta; p_t; m) = \frac{\cosh \eta}{\sqrt{\frac{  m^2  \,+\,p_t^2}{p_t^2}\,\,+\,\,\sinh^2 
\eta}}, \label{Jac}
\eeq
where $m$ is the typical mass of the produced particle, and $p_t$ is its typical transverse 
momentum.
Of course, to plot the distribution (\ref{finres}) as a function of pseudo-rapidity, one also 
has to express rapidity $y$ in terms of pseudo-rapidity $\eta$; 
this relation is given by 
\beq
y (\eta; p_t; m) = {1 \over 2}\  \ln\left[{{\sqrt{\frac{  m^2  \,+\,p_t^2}{p_t^2}\,\,+\,\,\sinh^2 \eta}} + \sinh \eta} \over 
{{\sqrt{\frac{  m^2  \,+\,p_t^2}{p_t^2}\,\,+\,\,\sinh^2 \eta}} - \sinh \eta}\right]; \label{yeta}
\eeq
obviously, $h(\eta; p_t; m) = {\partial y (\eta; p_t; m) / \partial \eta}$. 

We now have to make an assumption about the typical invariant mass $m$ of the gluon mini--jet. 
Let us estimate it by assuming that the slowest hadron in the mini-jet decay is 
the $\rho$-resonance, with energy $E_{\rho} = (m_{\rho}^2 + p_{\rho,t}^2 + p_{\rho,z}^2)^{1/2}$, 
where the $z$ axis is pointing along the mini-jet momentum.  
Let us also denote by $x_i$ the fractions of the gluon energy $q_0$ 
carried by other, fast, $i$ particles in the mini-jet decay. Since the sum of transverse (with respect 
to the mini-jet axis) momenta 
of mini-jet decay products is equal to zero, the 
mini-jet invariant mass $m$ is given by 
\begin{equation}
\label{decay}
m^2_{jet}\,\equiv\, m^2 = ( \sum_i x_i q_0 + E_{\rho})^2 - (\sum_i x_i
q_z + p_{\rho,z})^2 \,\simeq\,\,
2 \sum_i x_i  q_z \cdot ( m_{\rho,t} - p_{\rho,z}) \equiv 2 Q_s \cdot m_{eff},
\end{equation}
where $m_{\rho,t}=(m_{\rho}^2 + p_{\rho,t}^2)^{1/2}$.
In \eq{decay} we used that $\sum_i x_i =1$ and $q_0 \approx q_z = Q_s$.
Taking $p_{\rho,z} \approx p_{\rho,t}  \approx\,300$ MeV and $\rho$ mass, we obtain $m_{eff}
\approx 0.5\,$ GeV.

We thus use the mass $m^2 \simeq 2 Q_s m_{eff} \simeq Q_s \cdot 1$ GeV in Eqs.(\ref{Jac},\ref{yeta}).
 Since the typical transverse momentum of the produced 
gluon mini--jet is $Q_s$, we take $p_t = Q_s$ in (\ref{Jac}).
The effect of the 
transformation from rapidity to pseudo--rapidity is the decrease of multiplicity at 
small $\eta$ by 
about $25-30 \%$, leading to the appearance of the $\approx 10 \%$ dip in the pseudo--rapidity 
distribution in the 
vicinity of $\eta = 0$. 
 We have checked that the change in the value of the mini--jet mass by two times affects the 
Jacobian at central pseudo--rapidity to about $\simeq 10\%$, leading to $\sim 3\%$ effect on the 
final result.

\begin{figure}[htbp] 
\begin{center}
\epsfig{file=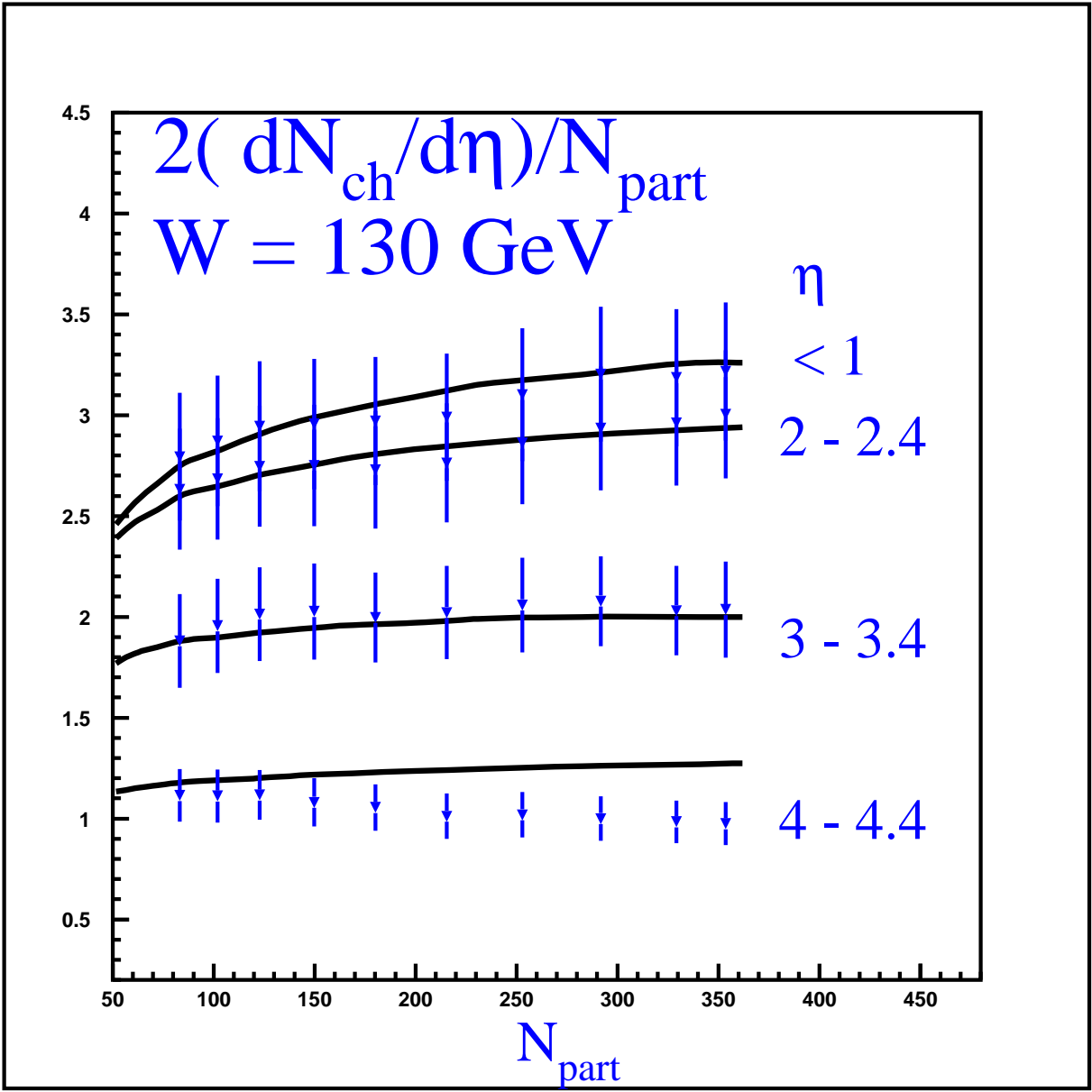,width=16cm}
\end{center}
\caption{ Centrality dependence of charged hadron production per participant at different 
pseudorapidity $\eta$ intervals in $Au-Au$ collisions 
at $\sqrt{s} = 130$ GeV; the data are from \cite{PHOBOS130}.}
\label{fig1}
\end{figure}
\begin{figure}[htbp] 
\begin{center}
\epsfig{file=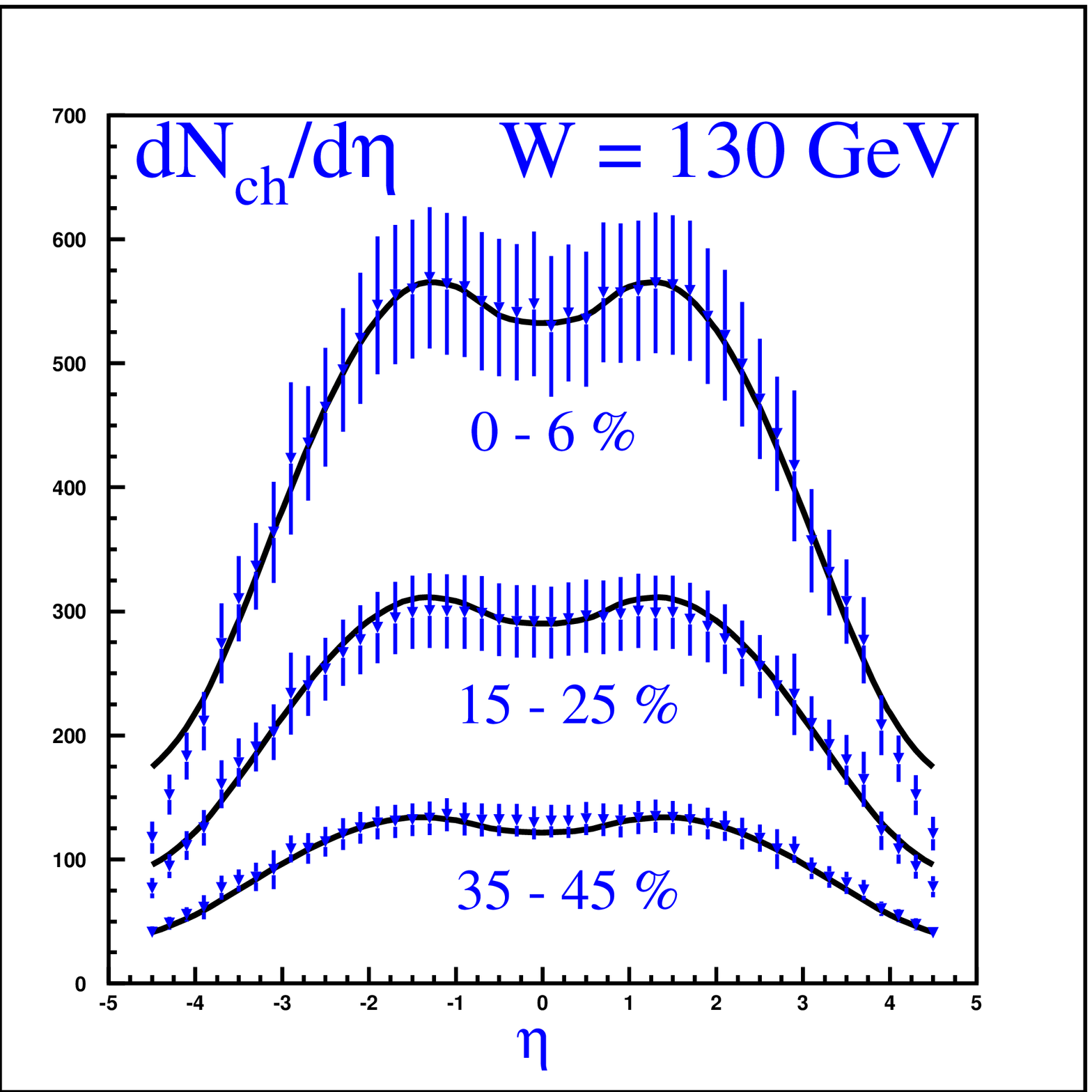,width=16cm}
\end{center}
\caption{ Pseudo--rapidity dependence of charged hadron production at different cuts on centrality 
in $Au-Au$ collisions 
at $\sqrt{s} = 130$ GeV; the data are from \cite{PHOBOS130}.}
\label{fig2}
\end{figure}
  \begin{figure}[htbp] 
\begin{center}
\epsfig{file=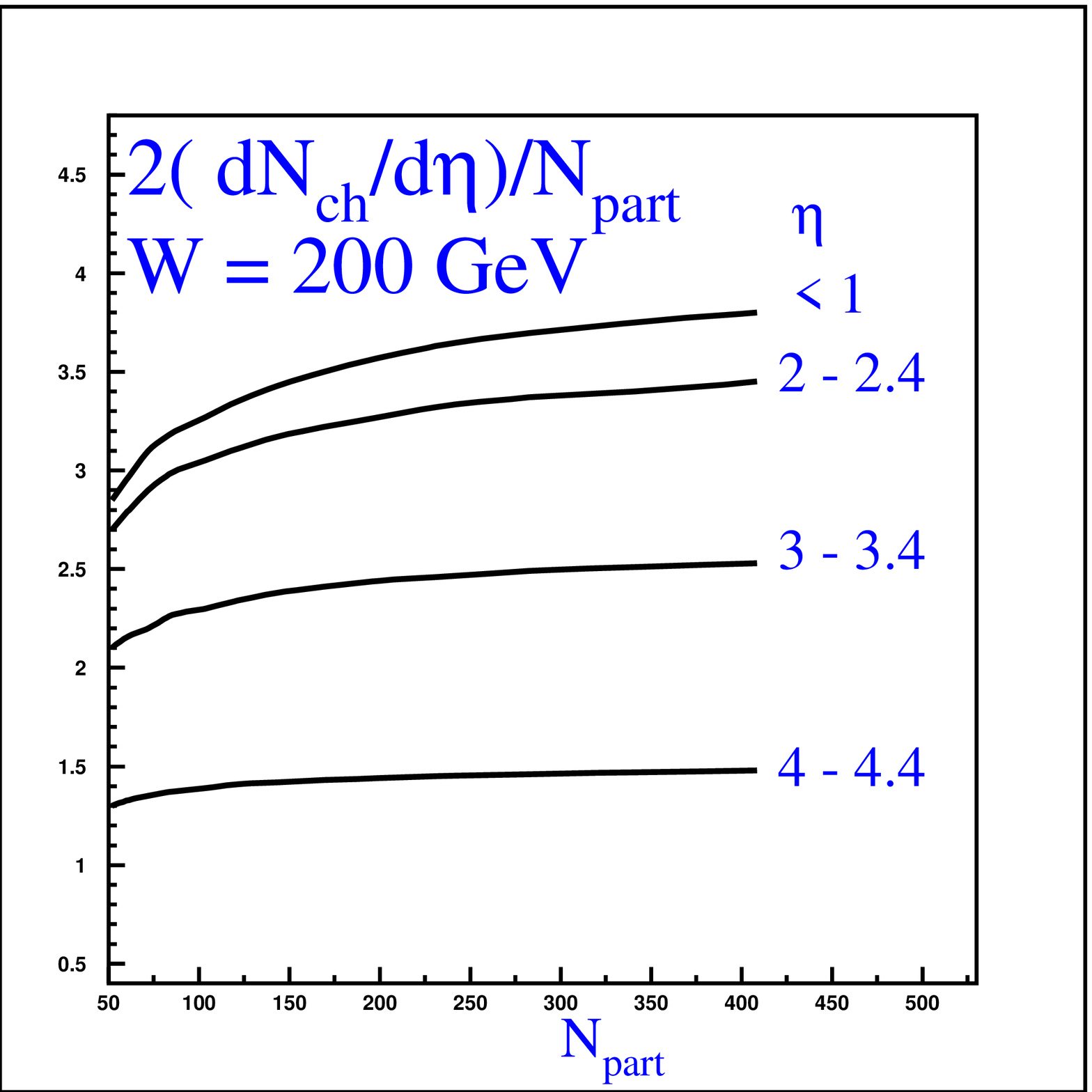,width=16cm }
\end{center}
\caption{Centrality dependence of charged hadron production per participant at different 
pseudorapidity $\eta$ intervals in $Au-Au$ collisions 
at $\sqrt{s} = 200$ GeV.}
\label{fig3}
\end{figure}
 
\begin{figure}[htbp] 
\begin{center}
\epsfig{file=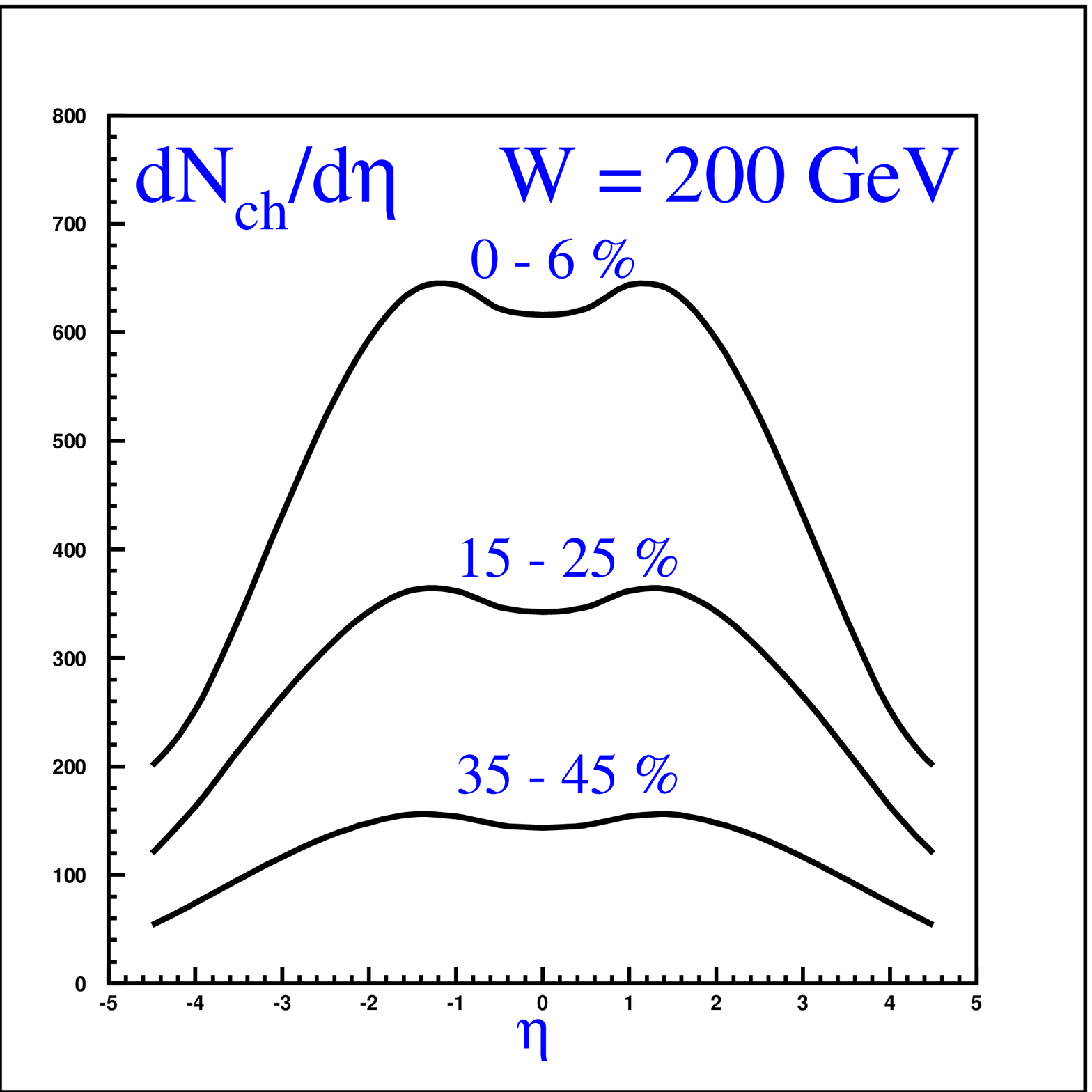,width=16cm}
\end{center}
\caption{Pseudo--rapidity dependence of charged hadron production at different cuts on centrality 
in $Au-Au$ collisions 
at $\sqrt{s} = 200$ GeV. }
\label{fig4}
\end{figure}

\vskip0.3cm

The results for the $Au-Au$ collisions at $\sqrt{s} = 130$ GeV are presented in Figs \ref{fig1} 
and \ref{fig2}. In the calculation, we use the results on the dependence 
of saturation scale on the mean number of participants at $\sqrt{s} = 130$ GeV 
from \cite{KN}, see Table 2 of that 
paper. The mean number of participants in a given centrality cut is taken from the PHOBOS paper 
\cite{PHOBOS130}. One can see that both the centrality dependence 
 and the rapidity dependence of the $\sqrt{s} = 130$ GeV PHOBOS data are well reproduced 
below $\eta \simeq \pm 4$. The rapidity dependence has been evaluated 
with $\lambda = 0.25$, which is within the 
range $\lambda = 0.25 \div 0.3$ inferred from the HERA data \cite{GW}. The discrepancy 
above $\eta \simeq \pm 4$ is not surprising since 
our approach does not properly take into account multi--parton correlations 
which are important in the fragmentation region.

Our predictions for $Au-Au$ collisions at $\sqrt{s} = 200$ GeV are presented in Figs. \ref{fig3} 
and \ref{fig4}. 
The only parameter which governs the energy dependence is the exponent $\lambda$, which we 
assume to be $\lambda \simeq 0.25$ as inferred from the HERA data. The absolute 
prediction for the multiplicity, as explained above, bears some uncertainty, but there is a definite 
feature of our scenario which is distinct from other approaches. 
It is the dependence of multiplicity on centrality, which around $\eta =0$ is determined 
solely by the running of the QCD strong coupling \cite{KN}. As a result, the centrality dependence 
at $\sqrt{s} = 200$ GeV is somewhat less steep than at $\sqrt{s} = 130$. While the 
difference in the shape at these two energies is quite small, in the perturbative 
mini-jet picture this slope 
should increase, reflecting the growth of the mini-jet cross section with 
energy. 

\begin{figure}[h]
\begin{minipage}{9.5cm}   
\begin{center}
\epsfysize=9.4cm
\leavevmode
\hbox{ \epsffile{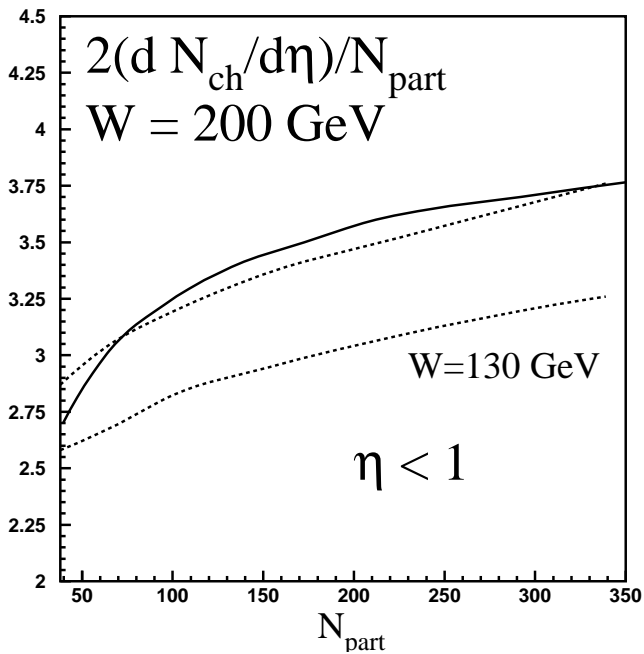}}
\end{center}
\end{minipage}
\begin{minipage}{5.5cm}
\caption{Comparison of the saturation (solid line) and ``soft plus hard'' (dashed line) 
calculations of the centrality 
dependence of charged hadron multiplicity at $\sqrt{s} = 200$ GeV; also shown is the 
``soft plus hard'' calculation at $\sqrt{s} = 130$ GeV (lower dashed curve). .}
\label{fig5}
\end{minipage}
\end{figure}

\vskip0.3cm

Let us confirm this statement quantitatively. 
The following simple parameterization \cite{KN} was found to describe the data \cite{data1,data2,data3} quite well:
\beq
{d N \over d \eta} = (1-X(s))\ n_{pp}\ {\langle N_{part} \rangle \over 2} + X(s)\  n_{pp}\ 
\langle N_{coll} \rangle, \label{soha}
\eeq
where $\langle N_{part(coll)} \rangle$ are the average numbers of participants (collisions) 
for a given centrality, and $X(s)$ is an $s-$ dependent parameter reflecting the relative 
strength of ``soft'' and ``hard'' components in multi-particle production. 
At $\sqrt{s} = 130$ GeV, the value deducted from the central multiplicity 
was $X(130 \ {\rm GeV}) = 
0.09 \pm 0.03$, and at $\sqrt{s} = 56$ GeV, it was $X(56 \ {\rm GeV}) = 
0.05 \pm 0.03$. In the perturbative picture, 
the ``hard'' component of multiplicity is proportional 
to the mini--jet production cross section. At $\eta = 0$, the energy dependence of the 
mini--jet cross section is determined by the {\it square} of the gluon structure function $x G(x)$  
(note that in the saturation approach, as we discussed above, it is the first power of $x G(x)$).
Taking $x G(x) \sim x^{-\lambda}$, with $\lambda \simeq 0.3$, translates then in the increase 
of the mini-jet production cross section by factor of $(130/56)^{2 \lambda} \simeq 1.7$ in going 
from  $\sqrt{s} = 56$ GeV to $\sqrt{s} = 130$ GeV, in accord with the ratio of the extracted values 
of $X(s)$ 
quoted above. Extrapolating the same energy dependence to  $\sqrt{s} = 200$ GeV, we estimate  
\beq
X(\sqrt{s} = 200\ {\rm GeV}) = \left({200 \over 130}\right)^{2 \lambda}\ X(\sqrt{s} = 130\ {\rm GeV}) 
\simeq 0.12 \pm 0.02,  \label{xs}
\eeq
where the error bar reflects only the uncertainty in the energy 
extrapolation of the central values of $X$. 
The value (\ref{xs}), together with Eq.(\ref{soha}), represents the prediction of the 
``soft plus hard'' approach for the centrality dependence at $\sqrt{s} = 200$ GeV.
Using the value (\ref{xs}) together with the average numbers of participants from \cite{KN}
(see \cite{KLNS} for details of the Glauber model we use) 
$\langle N_{part} \rangle = 339$ and collisions $\langle N_{coll} \rangle = 1049$ in the $6\%$ 
centrality cut at $\sqrt{s} = 200$ GeV, and the $pp$ multiplicity of $n_{pp} = 2.43$ which follows 
from the parameterization of the data $n_{pp} = 2.5 - 0.25 \ln(s) + 0.023 \ln^2(s)$ 
\cite{CDF,PHOBOS130},
we get the central value
\beq
{d N \over d \eta}(\eta = 0) \simeq 668, \label{shpred}
\eeq
which represents a $\simeq 20 \%$ increase compared to $\sqrt{s} = 130$ GeV.
This value is larger than the one predicted above on the basis of the 
initial--state saturation approach; 
however the main difference between the two scenarios will be in the centrality and 
rapidity dependences of hadron multiplicity. 

To estimate the difference in the shape 
of centrality dependence, let us consider the case when the multiplicities in central 
collisions are forced to be the same and equal to $d N / d \eta(\eta = 0) \simeq 634$, 
which is the upper limit of our prediction (\ref{pred200}) and below 
the central value (\ref{shpred}) (note that this represents the ``worst case'' in which 
the value of multiplicity in central collisions alone cannot be used to distinguish the two 
approaches). In the ``soft plus hard'' 
model, this multiplicity would then correspond to $X = 0.106$. In Fig. \ref{fig5}, we compare 
the resulting centrality dependences in the saturation and in the ``soft plus hard'' 
calculations. While the difference between the two curves is not large, the variation of 
the predictions with energy and 
rapidity could help to discriminate between the two approaches; for comparison, we also 
show the result of the ``soft plus hard'' calculation at $\sqrt{s} = 130$ GeV. 
  
\vskip0.3cm

To summarize, we have extended the previous 
analysis \cite{KN} of multi--particle production in the quasi--classical picture  
 to include 
the pseudo-rapidity dependence, and found good agreement with RHIC data at $\sqrt{s}=130$ GeV. 
If this agreement persists at higher energies, one may conclude that the dynamics 
of quasi--classical color fields, encoded in our approach in terms of a simple 
scaling function (\ref{finres}), 
describes well the gross features of multi-particle 
production already at RHIC energy.   
The experimental study of centrality and rapidity dependences at $\sqrt{s} = 200$ GeV 
will thus be a very important step in establishing the presence of high density QCD  
effects in relativistic nuclear collisions. The clear discrimination between this and other 
approaches may however require the study of more ``microscopic'' observables.
  
\vskip0.3cm

{\it Note added:} While we were finalizing this paper, the 
first $\sqrt{s}= 200$ GeV multiplicity measurement by the PHOBOS Collaboration 
had been announced \cite{ann}. 
The ratio of multiplicities in the central ($0-6 \%$ centrality cut) events at 
$\sqrt{s}= 200$ and $\sqrt{s}= 130$ GeV has been reported: $R(200/130) = 1.14 \pm 0.06$. 
While this is in accord with the prediction based on the saturation approach given above, 
$R(200/130) = 1.10 \div 1.14$, 
it is also consistent with the $20\%$ increase which we expect in 
the ``soft plus hard'' 
scenario. (We did not attempt to adjust to the  
central value of the announced data our calculations shown in 
Figs \ref{fig3} and \ref{fig4} and reported (e.g., \cite{DK1}) 
before the multiplicity in head--on collisions had been available). 
The crucial test of the quasi--classical approach will thus be provided by the centrality and 
pseudo--rapidity 
dependences of hadron multiplicities.

\vskip0.3cm

The work of D.K. was supported by the US Department of Energy 
(Contract \# DE-AC02-98CH10886). 
The research of E.L. was supported in part by the Israel Science 
Foundation, founded by the Israeli Academy of Science and Humanities, 
and BSF \# 9800276.


\newpage
  
\begin{thebibliography}{99}
\bibitem{GLR}
L. V. Gribov, E. M. Levin and M. G. Ryskin, Phys. Rep. {\bf 100} (1983) 1.
\bibitem{MUQI}
A. H. Mueller and J. Qiu,  Nucl. Phys. {\bf B 268} (1986) 427.
\bibitem{BM} J.-P. Blaizot and A.H. Mueller, Nucl. Phys. {\bf B 289} (1987) 847.
\bibitem{MV}
L. McLerran and R. Venugopalan,  Phys. Rev.  {\bf D 49} (1994)
2233, 3352; {\bf D 50} (1994) 2225, {\bf D 53} (1996) 458, {\bf D 59} (1999)
094002.
\bibitem{YM} E. Iancu and L. McLerran, Phys.Lett. B510 (2001) 145; 
Yu. Kovchegov, Phys. Rev. D54 (1996) 5463; hep-ph/0011252; 
A. Krasnitz and R. Venugopalan, Phys. Rev. Lett. 84 (2000) 4309; 
E. Levin and K. Tuchin, Nucl.Phys. B573 (2000) 833. 

\bibitem{KV}
Ia. Balitsky,  Nucl. Phys.  {\bf B 463}  (1996) 99;\\
Yu. Kovchegov,
 Phys. Rev. {\bf D 60} (2000)  034008.

\bibitem{KN} D. Kharzeev and M. Nardi, Phys. Lett. B507 (2001) 121.

\bibitem{inst} D. Kharzeev and E. Levin, Nucl. Phys. B578 (2000) 351; D. Kharzeev, 
Yu. Kovchegov and E. Levin, Nucl. Phys. A690 (2001) 621; hep-ph/0106248. 
\bibitem{inst1} E.V. Shuryak, Phys. Lett. B486 (2000) 378; hep-ph/0101269; M. Nowak, 
E. Shuryak and I. Zahed, Phys.Rev. D64 (2001) 034008. 

\bibitem{data1} B. Back et al., PHOBOS Coll., Phys. Rev. Lett. 85 (2000) 3100; 
nucl-ex/0105011; nucl-ex/0106006.

\bibitem{data2} 
K. Adcox et al., PHENIX Coll.,  Phys. Rev. Lett. 86 (2001) 3500; nucl-ex/0012008; 
nucl-ex/0104015.

\bibitem{data3} 
 C. Adler et al., STAR Coll., nucl-ex/0106004.

\bibitem{data4} 
F. Videbaek et al., BRAHMS Coll., http://www.rhic.bnl.gov/qm2001.

\bibitem{WG} X.-N. Wang and M. Gyulassy, Phys. Rev. Lett. {\bf 86} (2001) 3496.

\bibitem{EKRT}  K.J. Eskola, K. Kajantie, P.V. Ruuskanen, K. Tuominen, 
Nucl. Phys. {\bf B570} (2000) 379;  K.J. Eskola, K. Kajantie, K. Tuominen, 
Phys. Lett. {\bf B497} (2001) 39; hep-ph/0106330.

\bibitem{Alf} A. Capella and D. Sousa, Phys. Lett. {\bf B511} (2001) 185.

\bibitem{alt} S. Jeon and J. Kapusta, Phys. Rev. {\bf C63} (2001) 011901; N. Armesto, 
C. Pajares, D. Sousa, hep-ph/0104269;
D. E. Kahana and S. H. Kahana, Phys. Rev. {\bf C63} (2001) 031901;  
A. Accardi, hep-ph/0107301.

\bibitem{DK} D. Kharzeev, {\rm nucl-th/0107033}.

\bibitem{AHM} A.H. Mueller, Nucl. Phys. {\bf B572} (2000) 227.

\bibitem{Raju} A. Krasnitz and R. Venugopalan, Phys. Rev. Lett. {\bf 86} (2001) 1717.

\bibitem{Yuri} Yu. Kovchegov, hep-ph/0011252. 

\bibitem{GW} K. Golec-Biernat and M. W{\"u}sthof, Phys. Rev. {\bf D59} (1999) 014017; 
Phys. Rev. {\bf D60} (1999) 114023; 
A. Stasto, K. Golec-Biernat and J. Kwiecinski, Phys. Rev. Lett. {\bf 86} (2001) 596.  

\bibitem{GM} M. Gyulassy and L. McLerran, Phys. Rev. {\bf C56} (1997) 2219.

\bibitem{qcr} G. Farrar and S.J. Brodsky, Phys. Rev. Lett. {\bf 31} (1973) 1153;\\  
V.A. Matveev, R.M. Muradian, A.N. Tavkhelidze, Lett. Nuovo Cim. {\bf 7} (1973) 719. 

\bibitem{PHOBOS130} B. Back et al., PHOBOS Coll., nucl-ex/0106006.

\bibitem{KLNS} D. Kharzeev, C. Louren{\c{c}}o, M. Nardi and H. Satz, 
Z. Phys. {\bf C 74} (1997) 307.

\bibitem{CDF} F. Abe et al., CDF Coll., Phys. Rev. {\bf D41} (1990) 2330.

\bibitem{ann} B. Wyslouch, PHOBOS Coll., Talk at the INPC 2001, Berkeley, USA, July 30 - August 2, 2001. 

\bibitem{DK1} D. Kharzeev, Talk at the ``Glauber approach at RHIC'' Workshop, BNL, July 19, 2001; 
http://www.phobos.bnl.gov/GlauberWorkshop.htm



\end{thebibliography}
\end{document}